\documentclass[amsfonts,12pt]{article}
\usepackage{amsmath,amssymb,graphicx,makeidx}

\author{Lorenz Hartmann\\
{\small Ludwig-Maximilians-Universit\"{a}t M\"{u}nchen,}\\
{\small Theresienstr. 37, 80333 M\"{u}nchen}\\
John R. Klauder\\
{\small Departments of Physics and Mathematics, University of Florida,}\\
{\small Gainesville, FL 32611}}
\title{Weak Coherent State Path Integrals}

\begin{document}
\date{}
\maketitle

\begin{abstract}

Weak coherent states share many properties of the usual coherent states, 
but do not admit a resolution of unity expressed in terms of a local integral. 
They arise e.g. in the
case that a group acts on an inadmissible fiducial vector. Motivated
by the recent Affine Quantum Gravity Program, the present article
studies the path integral 
representation of the affine weak coherent state matrix elements
of the unitary time-evolution operator. 
Since weak coherent states do not admit a resolution of unity, it is clear that
the standard way of constructing a path integral, by time slicing, is 
predestined to fail. Instead a well-defined path integral with Wiener measure,
based on a continuous-time regularization, is
used to approach this problem. The dynamics is rigorously established for
linear Hamiltonians, and the difficulties presented by 
more general Hamiltonians are 
addressed.  

\end{abstract}

\section{Introduction}\label{Introduction}

Unlike the standard phase space path integrals constructed by the time slicing 
method, 
the path integral with Wiener measure invented by Klauder, Daubechies and others
uses a continuous-time regularization 
factor \cite{DK}\cite{DKP}\cite{KlauderQiG}\cite{KlauderUQ}\cite{KlauderWCS}. 
This path integral is
$\pmb{\int}\exp\{-i{\textstyle \int}[qdp+dG(p,q)+h(p,q)dt]\}
\:d\mu^\nu_W$, where $G$ is an arbitrary $C^1$ function and $h$ is the
classical Hamiltonian in a sense which will be explained later. The pinned
Wiener measure $d\mu_W^\nu$ is defined with the help of the heat kernel 
$\int d\mu^\nu_W:=
[\exp\{\nu T\Delta_{LB}\}](p^{\prime\prime}, q^{\prime\prime}, q^\prime, 
p^\prime)$. Thus, by way of the Laplace-Beltrami operator $\Delta_{LB}$,
a metric
is introduced. The formal phase space path integral
${\cal N}_\nu\pmb{\int}\exp\{-i{\textstyle \int}
[q\dot{p}+\dot{G}(p,q)+h(p,q)]dt\}\exp\{-{\textstyle \frac{1}{2\nu}}
{\textstyle \int}(d\sigma^2/dt^2)dt\}{\cal D}p{\cal D}q$ 
can be given meaning by equating it to the above Wiener measure
path integral. Here, ${\cal N}_\nu$ is a formal normalization constant, and
$d\sigma^2$ is the metric mentioned above.  
The variables $p$ and $q$ in the well-defined Wiener measure path integral 
are stochastic variables describing 
Brownian Bridges. The integral $\int qdp$ has to be 
interpreted
as a stochastic integral. The rule adopted here is the Stratonovich mid-point
rule $\int qdp:=\lim\sum\frac{1}{2}(q_{l+1}+q_l)(p_{l+1}-p_l)$, which guarantees
that the ordinary rules of calculus still apply. It was shown, first for
the case of a flat and spherical phase space metric \cite{DK}, then for a
hyperbolic metric \cite{DKP}, that the limit of diverging diffusion constant
$\nu$ exists for a wide set of quantum Hamiltonians ${\cal H}$, including
at least all Hamiltonians polynomial in the basic quantum kinematical operators. 
The limit is equal to the coherent state matrix element
$\langle p^{\prime\prime}q^{\prime\prime}|\exp\{-iT{\cal H}\}|p^\prime 
q^\prime\rangle$ of the unitary time-evolution operator and the specific metric 
determines the coherent states in question: 
The flat metric is inevitably connected 
with the coherent states of the Heisenberg-Weyl group (and in the canonical,
Cartesian form, it is connected to the canonical coherent states), the 
spherical metric is associated with the coherent states of the SU(2) group, and
the hyperbolic metric leads to the coherent states of the affine group. And
with each group comes a set of quantum kinematical operators. Thus,
one can say that in these three cases the choice of geometry augmenting the 
classical phase space
manifold determines the quantum kinematical operators uniquely!
Furthermore, the classical Hamiltonian that goes with the quantum Hamiltonian
${\cal H}$ is given by the lower symbol\footnote{Other authors
call this symbol the upper symbol, since it is involved in an upper bound in
the Berezin-Lieb inequalities.} $h$, 
implicitly defined by the relation 
${\cal H}=\int h(p,q)|pq\rangle\langle pq|d\mu(p,q)$. Here,
$d\mu(p,q)$ is the left-invariant group measure of the group which defines 
the coherent states. This measure is normalized such that $h(p,q)\equiv1$
leads to ${\cal H}=1\!\!1$, and, thus, provides the usual resolution of unity. 
Since the Stratonovich rule
is used, and since the coherent states merely change labels under canonical
(coordinate) transformations, apart from possible phase factors, the path 
integral
$\langle p^{\prime\prime}q^{\prime\prime}|\exp\{-iT{\cal H}\}|p^\prime 
q^\prime\rangle=\lim_{\nu\rightarrow\infty}\pmb{\int}\exp\{-i{\textstyle \int}
[qdp+dG(p,q)+h(p,q)dt]\}\:d\mu^\nu_W$ is covariant under canonical (coordinate)
transformations and the quantization is fully geometric in 
nature \cite{KlauderQiG}\cite{KlauderUQ}. The foregoing has been extended to 
arbitrary geometries of the phase space \cite{Maraner}.

In an attempt to quantize gravity \cite{KlauderAQG1}\cite{KlauderAQG2}, 
Klauder was led to consider
affine rather than canonical commutation relations for the field operators (the
spatial part of the metric and its partner field). In the simplest case of
constant fields, the problem reduces to a toy model of just one degree of
freedom, namely the affine coherent states. To be more precise, it includes
the affine coherent states, which fulfill a fiducial vector admissibility 
condition \cite{DKP}\cite{KlauderWCS}\cite{Aslaksen}, but also those states 
which violate it. 
These latter states 
do not resolve unity anymore and, therefore, are called weak coherent
states. The Affine Quantum Gravity Program has provided the motivation to raise
the question of the existence of path integrals for these weak coherent states.

It is clear that a path integral can not be constructed with weak coherent
states in the standard way, since the resolution of unity is the key to the
time-slicing approximation. However, the extension of the well-defined
path integral with Wiener measure introduced above to the situation of 
weak coherent states could still be possible, and this is the goal of the 
present article.
Two different methods to extend the Wiener measure path
integral will be introduced: the first is based on the spectral decomposition 
of certain 
operators and will therefore be called the ``spectral approach". Unfortunately,
it is limited to one very special case. The second uses an extra regularization
parameter and is consequently called the ``regularizing approach". In both 
cases, the path integral for zero Hamiltonian is studied first, while the 
dynamics is introduced as a second step.

\section{Weak coherent state path integrals}

\subsubsection*{General definitions}

Coherent states are defined by two properties \cite{Klauder}:\\

{\bf 1) Continuity}: The states $|l\rangle$ are a strongly continuous 
vector-valued function of the label $l$.\\

{\bf 2) Resolution of unity}: There exists 
a positive measure $\delta l$ on the label 
space $\mathfrak{L}$ such that the identity operator $1\!\!1$
on $\mathfrak{H}$ can, upon integration over $\mathfrak{L}$, be represented as 

\begin{equation*}\label{ResUnity}
1\!\!1=\int |l\rangle\langle l|\delta l
\end{equation*}

A more general class of states can be obtained by relaxing the second 
property:\\

{\bf $\mathbf2^\prime$) Completeness}: 
The family of vectors $(|l\rangle)$ is total, i.e.,
the closed linear span of $(|l\rangle)$ is the whole Hilbert space 
$\mathfrak{H}$.\\

\noindent States which share the properties $1)$ and $2^\prime )$ 
have been named 
Klauder states 
\cite{Hartmann}. They are the disjoint union of the coherent states in the
sense above and the weak coherent states, which do not possess a resolution of
unity.

\subsubsection*{Affine weak coherent states}

The affine group $(M_+,\circ)$ is the set $M_+:=\mathbb{R}^+\times\mathbb{R}$
with the group law $(q,p)\circ(q^\prime,p^\prime)=
(qq^\prime, p+q^{-1}p^\prime)$ and has two nontrivial, inequivalent, 
irreducible,
unitary representations \cite{DKP} $U_\pm(p,q)=e^{\pm ipQ}e^{-i\ln qD}$, where
the generators $Q>0$ and $D$ obey the affine commutation relation $[Q,D]=iQ$. 
The uncertainty product of the irreducible, self-adjoint operators $Q$ and $D$ is
$\Delta Q\Delta D\geq\frac{1}{2}\langle Q\rangle$. Setting $\langle Q\rangle
=1$ leads to a one-parameter family of minimum uncertainty states given in
$x$-representation by \cite{KlauderWCS} 
$\eta_\beta(x)=N_\beta x^{\beta-1/2}e^{-\beta x}$ with
normalization $N_\beta=(2\beta)^\beta\Gamma^{-1/2}(2\beta)$.
The affine coherent states are defined as 
$|pq\rangle:=U_+(p,q)|\eta_\beta\rangle$.
The group acts on admissible fiducial vectors, which 
fulfill \cite{DKP}\cite{KlauderWCS}\cite{Aslaksen}
$\langle Q^{-1}\rangle=\int_0^\infty x^{-1}|\eta_\beta(x)|^2dx<\infty$. Namely
these are the states with $\beta>1/2$. Weak coherent states, on the other hand,
are generated by the same group action on fiducial vectors with 
$0<\beta\leq1/2$. 
For the whole parameter range $0<\beta$, the overlap reads

\begin{equation*}\label{AffineOverlap}
\langle pq|rs\rangle=(qs)^{-\beta}2^{2\beta}[(q^{-1}+s^{-1})+i\beta^{-1}
(p-r)]^{-2\beta}
\end{equation*}

The construction of the affine coherent state path integral with Wiener 
measure \cite{DKP} is based on a linear complex polarization condition. 
For the minimum uncertainty fiducial vectors, 
$(Q-1+i\beta^{-1}D)|\eta_\beta\rangle=0$ holds. Hence, 
all functions $\psi(p,q):=
\langle pq|\psi\rangle$ are annihilated by the operator $B=-iq^{-1}\partial_p
+1+\beta^{-1}q\partial_q$. The same is true for the second-order differential
operator 

\begin{equation*}\label{OperatorA}
A:={\textstyle\frac{1}{2}}\beta B^\dagger B
={\textstyle\frac{1}{2}}
\{-\beta^{-1}\partial_qq^2\partial_q-\beta q^{-2}\partial^2_p-1+\beta
-2i\beta q^{-1}\partial_p\}
\end{equation*}

\noindent which is a nonnegative, self-adjoint operator with spectrum

\begin{equation*}\label{SpecA}
spec(A)=\{(\beta-{\textstyle\frac{1}{2}})^2-(\beta-{\textstyle\frac{1}{2}}-n)^2;
n\in\mathbb{N}, n<\beta-{\textstyle\frac{1}{2}}\}
\cup [(\beta-{\textstyle\frac{1}{2}})^2,\infty)
\end{equation*}

For $\beta>1/2$, the operator $A$ has a discrete eigenvalue $0$ and it follows, 
for $T>0$, that $\lim_{\nu\rightarrow\infty}[e^{-\nu TA}]\delta(p-p^\prime)
\delta(q-q^\prime)|_{p=p^{\prime\prime},q=q^{\prime\prime}}=[P_0]
(p^{\prime\prime},q^{\prime\prime};p^\prime,q^\prime)$, 
where the expression on the
right hand side is the kernel of the projection operator onto the ground state.
But this kernel is also given by $(2\pi)^{-1}(1-\frac{1}{2\beta})\langle
p^{\prime\prime} q^{\prime\prime}|p^\prime q^\prime\rangle$. 
This is the key part of the construction, since
the rest follows by the Feynman-Kac-Stratonovich representation of the
kernel of $e^{-\nu TA}$, which is ${\cal N}_\nu{\pmb\int}e^{-i{\textstyle\int}
q\dot{p}dt-{\textstyle\frac{1}{2\nu}\int}[\beta^{-1}q^2\dot{p}^2+\beta q^{-2}
\dot{q}^2]dt}{\cal D}p{\cal D}q$. As stated in the introduction, this formal
expression makes sense as a Wiener measure path integral, and so finally

\begin{equation*}
\langle p^{\prime\prime} q^{\prime\prime}|p^\prime q^\prime\rangle=
\lim_{\nu\rightarrow\infty}2\pi(1-{\textstyle\frac{1}
{2\beta}})^{-1}e^{\nu T/2}{\pmb\int}e^{-i{\textstyle\int}q\,dp}\:d\mu_W^\nu(p,q)
\end{equation*}

\noindent which is a well-defined expression\footnote{The path 
integral for a 
non-zero Hamiltonian is constructed in much the same way. The only
difference is that $\nu A$ must be replaced by an operator involving
the Hamiltonian $h$, namely $\nu A+ih$.}.  

For $0<\beta\leq1/2$, i.e., in the weak coherent state case, 
the operator $A$ has only a continuous spectrum, and the 
limit of diverging diffusion constant of the operator $e^{-\nu TA}$ is zero. 
Thus, the whole construction outlined above breaks down. To prevent this
collapse to a trivial result, two different approaches will be discussed.

\subsection{Spectral approach}

The idea in this approach is to determine a $\nu$-dependent rescaling factor,
such that the limit of diverging diffusion constant will be nontrivial. This 
was proposed by Klauder \cite{KlauderWCS}.

\subsubsection*{The general case}

Let X be a non-negative self-adjoint operator on a certain Hilbert space and
assume zero is in its continuous, but not in its discrete, spectrum. 
The operator $X$ generates
a semigroup $e^{-\nu TX}$, which has a spectral representation
$e^{-\nu XT}=\int_0^\infty e^{-\nu\lambda T}d\mathbb{ E}(\lambda)$ or
$\langle x^{\prime\prime}|e^{-\nu TX}|x^\prime\rangle=
\int_0^\infty e^{-\nu\lambda T}d\langle x^{\prime\prime}|\mathbb{ E}(\lambda)|
x^\prime\rangle$. 

Since only well-behaved potentials will eventually be of interest, the
reasonable assumption is made that the measure
$d\langle x^{\prime\prime}|\mathbb{ E}(\lambda)|
x^\prime\rangle$ has an absolutely continuous, but no singularly continuous part.
Then the spectral family can be written as a (weighted) integral over 
one-dimensional projection operators 
$\mathbb{ E}(\lambda)=\int_{-\infty}^\lambda|E\rangle\langle E| 
\rho(E)dE$ \footnote{For a singularly continuous measure this would not be
possible: $\mu_{sc}(x)=\int_{-\infty}^x d\mu_{sc}(y)\neq
\int_{-\infty}^x(d\mu_{sc}/dy)dy=0$ since $d\mu_{sc}/dy=0$ almost everywhere.}.
If the generalized eigenstates $|E\rangle$ are $\delta$-orthonormalized, then
$\rho(E)=1$. 

The matrix element of $e^{-\nu TX}$ can then be written as

\begin{equation}\label{OperatorX}
\langle x^{\prime\prime}|e^{-\nu TX}|x^\prime\rangle=
\int_0^\infty e^{-\nu\lambda T}\psi_\lambda(x^{\prime\prime})
\psi^*_\lambda(x^\prime)\rho(\lambda)d\lambda
\end{equation}

\noindent and the $\psi_\lambda$ are
continuous in $\lambda$. Moreover, $\rho$ - being part of the measure - is
at least right-continuous. For $\delta$-orthonormalized wavefunctions,
$\rho(\lambda)\equiv 1$. 

The goal is to find the rescaling factor which saves Eq. 
(\ref{OperatorX}) from becoming trivial in the limit of diverging diffusion
constant $\nu$. Since, for very large $\nu$, the factor $e^{-\nu T\lambda}$
suppresses everything but the values for very small $\lambda$, the behavior of 
$f_{x^\prime, x^{\prime\prime}}(\lambda):=\psi_\lambda(x^{\prime\prime})
\psi^*_\lambda(x^\prime)\rho(\lambda)$ near $\lambda=0$ is all that matters. 
To give an example, assume that $f_{x^\prime,x^{\prime\prime}}(\lambda)
\propto \lambda^a$ for small $\lambda$. 
Now, the proper rescaling factor can be determined,
and in the example it is
 
\begin{equation}\label{Example}
\int_0^\infty d\lambda\, \lambda^a e^{-\nu\lambda T}=
\frac{\Gamma(a+1)}{(\nu T)^{a+1}}
\end{equation}

\noindent 
After rescaling with the inverse one gets 
$\frac{(\nu T)^{a+1}}{\Gamma(a+1)}\lambda^a
e^{-\nu\lambda T}\stackrel{\nu\rightarrow\infty}{\longrightarrow}
\delta(\lambda)$ which represents a $\delta$-function weight on $\lambda=0$. 

The rescaling factor can be computed self-consistently,
and the general formula reads

\begin{equation}\label{GeneralSpectralApproach}
\frac{\int_0^\infty e^{-\nu\lambda T}
\psi_\lambda(x^{\prime\prime})\psi^*_\lambda(x^\prime)
\rho(\lambda)d\lambda}{\int_0^\infty e^{-\nu\lambda T}
\psi_\lambda(0)\psi^*_\lambda(0)
\rho(\lambda)d\lambda}\stackrel{\nu\rightarrow\infty}{-\!\!\!\rightharpoonup}
\frac{\psi_0(x^{\prime\prime})
\psi^*_0(x^\prime)}{\psi_0(0)
\psi^*_0(0)}
\end{equation}

\noindent The numerator of the last expression, $\psi_0(x^{\prime\prime})
\psi^*_0(x^\prime)$, is the kernel of the desired projection 
operator onto the ground state, and we have assumed that the denominator is nonzero.  The convergence is in a distributional sense
(denoted by the symbol $\rightharpoonup$). If the functional form of
$\psi_0(x^{\prime\prime})\psi^*_0(x^\prime)$ is known to be continuous, then
the convergence is pointwise. 

Observe, in the example with $f_{x^\prime,x^{\prime\prime}}(\lambda)=\lambda^a$,
one must have $a>-1$, or else the rescaling factor would be identically zero 
(since the integral would be infinity). But, since the rescaling factor can
be determined self-consistently, i.e., by the denominator of Eq.
(\ref{GeneralSpectralApproach}), which always exists, there is no hidden 
``trap" to look out for. Moreover, the evaluation of the denominator
need not necessarily be at the point $x^{\prime\prime}=x^\prime=0$. It
could be at any point $x^{\prime\prime}=x^\prime=b$, $b\in\mathbb{R}$, or even
$b=\pm\infty$, as long as the function $\psi_\lambda(x)$ is not $0$ at $b$.  
Whatever gives the easiest result is the preferred choice. And the arbitrariness
of this choice is not critical: Assume $K$ to be the reproducing kernel
of some reproducing kernel Hilbert space, and let $a$ be a positive constant. 
Then,
$aK$ is just as good a reproducing kernel, since the same class of functions
arises, only the inner product has to
be redefined. 

\subsubsection*{The affine case}

The foregoing is now applied to the case of the affine weak coherent states. 
Unfortunately,
$A$ matches the required properties, namely that $0$ be in the continuous 
spectrum, only in the case $\beta=1/2$! This is true in spite of the fact
that $A\langle pq|\psi\rangle=0$ (for arbitrary $|\psi\rangle$), since
an equation $A\psi=\alpha\psi$ need not necessarily imply $\alpha\in spec(A)$. 
In fact, the $\psi(p,q)=\langle pq|\psi\rangle$ are not generalized eigenvectors
except in the case $\beta=1/2$ \cite{Hartmann}. Consequently, the isolating
procedure can only be performed for $\beta=1/2$, and the general theory above
ensures the existence of the weak coherent state path integral. 

For the case at hand a connection between the operator $A$ and the 
one-dimensional Morse operator $H_{Morse}$ exists \cite{DKP} and makes 
the explicit functional form of the generalized eigenfunctions available. With
the aid of these, the rescaling factor can be computed explicitly. 

The problem to find the eigenfunctions of the operator $A$ 
is first reduced to a problem on 
$L^2(\mathbb{ R}^+)$ and then to a problem on $L^2(\mathbb{ R})$, leading 
to the Morse operator:

\begin{eqnarray}\label{ConnectionAH}
& & A\langle U(p,q)\phi|\psi\rangle=\langle A^*U(p,q)\phi|\psi\rangle\nonumber\\
& = & {\textstyle\frac{1}{2}}\langle [-\beta^{-1}\partial_q q^2\partial_q-\beta 
q^{-2}\partial^2_p
-2i\beta q^{-1}\partial_p+\beta-1]
e^{ipQ}e^{-i\ln qD}\phi|\psi\rangle\nonumber\\
&=&{\textstyle\frac{1}{2}}\beta^{-1}\langle e^{ipQ}e^{-i\ln qD}
\{D^2+iD+\beta^2Q^2\
-2\beta^2Q+\beta^2-\beta\}(Q^{1/2}\phi^\prime)
|\psi\rangle\nonumber\\
&=&{\textstyle\frac{1}{2}}\beta^{-1}\langle e^{ipQ}e^{-i\ln qD}Q^{1/2}\{D^2
+\beta^2Q^2\
-2\beta^2Q+(\beta-1/2)^2\} \phi^\prime|\psi\rangle
\end{eqnarray}
where $\phi=Q^{1/2}\phi'$. 

Under the unitary transformation 

\begin{equation}\label{U}
(\tilde{U}\psi)(x)=e^{x/2}\psi(e^x)
\end{equation}

\noindent the operator in
braces in the last line of 
Eq. (\ref{ConnectionAH}) (called $H$ in \cite{DKP}) 
is transformed to the Morse operator:

\begin{equation}
H_{Morse}=-{\textstyle\frac{d^2}{dx^2}}
+\beta^2(e^{2x}-2e^x)+(\beta-{\textstyle\frac{1}{2}})^2
\end{equation}

The eigenfunctions of the Morse operator can be found in \cite{Grosche}, and,
for $\beta=1/2$, they are given in momentum representation (and 
$\delta$-orthonormalized) by

\begin{equation}\label{MorseWavefunctionsLambda}
\psi_\lambda(x)
=({\textstyle\frac{\lambda\sinh(2\pi\lambda)}
{\pi^2}})^{1/2}\Gamma(i\lambda)e^{-x/2}W_{1/2,i\lambda}
(e^x)
\end{equation}

\noindent where $W$ is a Whittaker function. With a mass $m=1/2$, one has
the relation $E=\lambda^2$ for energy and momentum, 
and the $\delta$-orthonormalized eigenfunctions
in energy representation are

\begin{equation}\label{MorseWavefunctions}
\psi_E(x)=({\textstyle\frac{\sinh(2\pi\sqrt{E})}
{2\pi^2}})^{1/2}\Gamma(i\sqrt{E})e^{-x/2}W_{1/2,i\sqrt{E}}
(e^x)
\end{equation}

Since the Whittaker function $W_{1/2,0}(z)=e^{-z/2}z^{1/2}$, 
the $x$-dependence of
$\psi_{E=0}(x)$ is $e^{-e^x/2}$. Thus, the rescaling factor can best be 
determined with the choice $x^{\prime\prime}=x^\prime=b=-\infty$ where this
function is equal to one. For small $E$, the function $f_{-\infty,-\infty}(E)
=\psi_E(-\infty)\psi_E^*(-\infty)\rho(E)\approx \pi^{-1}E^{-1/2}$ because
$\sinh(2\pi\sqrt{E})\approx 2\pi\sqrt{E}$, $|\Gamma(i\sqrt{E})|^2\approx
1/E$. Inserting this $E$-dependence into the general formula ($\rho(E)=1$ 
because of $\delta$-orthonormalization),
one finds the inverse rescaling factor

\begin{equation}\label{InvRescalingFactor}
\int_0^\infty e^{-\nu TE}f_{-\infty,-\infty}(E)dE=(\pi\nu T)^{-1/2} 
\end{equation}

Because of the connection between the ``Morse"-level and the original 
problem [Eqs. (\ref{ConnectionAH}) and (\ref{U})], this is already the
proper rescaling factor for the original problem as well.

The sought-for weak coherent state path integral
for $\beta=1/2$ and vanishing Hamiltonian is thus

\begin{equation}\label{WCSPI}
\langle p^{\prime\prime}q^{\prime\prime}|p^\prime q^\prime\rangle
=\lim_{\nu\rightarrow\infty}K_\nu {\pmb \int}e^{-i
{\textstyle \int}q\,dp}\:d\mu_W^\nu
\end{equation}

\noindent with rescaling factor $K_\nu=(\pi\nu T)^{1/2}$.

\subsection*{Introducing dynamics}

Since the only case in which the spectral approach worked was $\beta=1/2$, this
value is assumed throughout the remainder of this subsection.
Dynamics are introduced by the quantum Hamiltonian ${\cal H}$, 
which is a function
of the basic kinematical operators $Q$ and $D$. The goal is to represent the
propagator $\langle p^{\prime\prime}q^{\prime\prime}|\exp\{-iT{\cal H}\}|
p^\prime q^\prime\rangle$ as a (weak coherent state) path integral. 
The expression

\begin{eqnarray}
&&\langle p^{\prime\prime}q^{\prime\prime}|e^{-iT{\cal H}}|
p^\prime q^\prime\rangle \nonumber\\
&=&\lim_{\nu\rightarrow\infty}K_\nu {\cal N}_\nu {\pmb \int}e^{-i
{\textstyle \int}[q\dot{p}+h_w(p,q)]dt}e^{-\frac{1}{2\nu}{\textstyle \int}
[\beta^{-1}q^2\dot{p}^2+\beta q^{-2}\dot{q}^2]dt}{\cal D}p{\cal D}q\nonumber\\
&=&\lim_{\nu\rightarrow\infty}K_\nu {\pmb \int}e^{-i
{\textstyle \int}[q\,dp+h_w(p,q)dt]}\:d\mu_W^\nu
\end{eqnarray}

\noindent was proposed \cite{KlauderWCS} as the path integral for a class of 
Hamiltonians which contains 
at least all Hamiltonians polynomial in $Q$ and $D$. The new symbol $h_w(p,q)$, 
interpreted as the classical Hamiltonian 
associated with the quantum Hamiltonian, is implicitly given by

\begin{equation}
\langle p^{\prime\prime}q^{\prime\prime}|{\cal H}|
p^\prime q^\prime\rangle
=\lim_{\nu\rightarrow\infty}K_\nu {\pmb \int}e^{-i
{\textstyle \int}q\,dp}\,[T^{-1}{\textstyle\int}h_w(p,q)dt]\:d\mu_W^\nu
\end{equation}

\noindent and will be called the weak symbol. 

The whole conjecture is based on the observation that, for a linear
Hamiltonian $RQ+SD$, the propagator can be reduced to a mere 
overlap \cite{KlauderWCS}\cite{Hartmann}: 

\begin{equation}
\langle p^{\prime\prime}q^{\prime\prime}|e^{-i(RQ+SD)T}|p^\prime q^\prime
\rangle
=\langle p^{\prime\prime}e^{ST}+R/S\cdot(e^{ST}-1), q^{\prime\prime}
e^{-ST}|p^\prime q^\prime\rangle
\end{equation}

Consequently, the problem is already solved for a linear Hamiltonian,
and what remains is to 
determine the weak symbol associated with ${\cal H}=RQ+SD$. 
According to Eq. (\ref{WCSPI}) the path integral for this Hamiltonian is

\begin{equation}
\lim_{\nu\rightarrow\infty}K_\nu {\pmb \int}^{p^{\prime\prime}e^{ST}
+R/S\cdot(e^{ST}-1), q^{\prime\prime}e^{-ST}}_{p^\prime, q^\prime}
e^{-i{\textstyle \int}q\,dp}\:d\mu_W^\nu
\end{equation}

Since this is a well-defined functional integral, one can change
integration variables

\begin{eqnarray*}
&&p(t)\rightarrow p(t)e^{St}+R/S(e^{St}-1)\\
&&q(t)\rightarrow q(t)e^{-St}
\end{eqnarray*}

\noindent and obtain $\exp\{-i\int(qe^{-St}
d[pe^{St}+R/S(e^{St}-1)]\}=\exp\{-i\int[q\,dp+(Rq+Spq)dt]\}$ as the
new integrand. The new measure
is\footnote{$(...)^\bullet$ means the time derivative of the expression
in parentheses}

\begin{eqnarray*}
d\tilde{\mu}_W^\nu&=&{\cal N}_\nu \exp\{-{\textstyle \frac{1}{2\nu}}
{\textstyle \int}
[\beta^{-1}(qe^{-St})^2(pe^{St}+{\textstyle \frac{R}{S}}(e^{St}-1))^{\bullet 2}
\nonumber\\
&&+\beta(qe^{-St})^{-2}(qe^{-St})^{\bullet 2}]dt\}
{\cal D}[pe^{St}+{\textstyle \frac{R}{S}}(e^{St}-1)]{\cal D}(qe^{-St})
\end{eqnarray*}

\noindent But, since the measure is actually

\begin{eqnarray*}
{\cal D}[pe^{St}+{\textstyle \frac{R}{S}}(e^{St}-1)]&=&
\lim_{\epsilon\rightarrow0}{\textstyle \prod}_{k=1}^N d[p(t)e^{St}
|_{t=k\epsilon}+{\textstyle \frac{R}{S}}(e^{St}-1)|_{t=k\epsilon}]\\
&=&\lim_{\epsilon\rightarrow0}{\textstyle \prod}_{k=1}^N[dp(t)
e^{St}|_{t=k\epsilon}+(pSe^{St}+Re^{St})dt|_{t=k\epsilon}]\\
&=&\lim_{\epsilon\rightarrow0}{\textstyle \prod}_{k=1}^N[dp_k
e^{Sk\epsilon}+(p_kSe^{Sk\epsilon}+Re^{Sk\epsilon})\epsilon]\\
&=&\lim_{\epsilon\rightarrow0}{\textstyle \prod}_{k=1}^N dp_k
e^{Sk\epsilon}={\cal D}p{\textstyle\prod}_te^{St}
\end{eqnarray*}

\noindent and analogous ${\cal D}(qe^{-St})={\cal D}q\prod_t e^{-St}$, 
the new measure can be expressed in terms of the old one as

\begin{eqnarray*}
d\tilde{\mu}_W^\nu&=&e^{-{\textstyle \frac{1}{2\nu}}{\textstyle \int}
[\beta^{-1}q^2((Sp+R)^2+2(Sp+R)\dot{p})+\beta q^{-2}(S^2q^2-2Sq\dot{q})]dt}
\:d\mu_W^\nu\\
&=&e^{-{\textstyle \frac{1}{2\nu}}{\textstyle \int}
[\beta^{-1}q^2((Sp+R)^2dt+2(Sp+R)dp)+\beta q^{-2}(S^2q^2dt-2Sqdq)]}
\:d\mu_W^\nu
\end{eqnarray*}

\noindent The first equality is again formal and gains meaning by the second
line,
where the stochastic integrals are understood in the Stratonovich sense, 
as usual. The change of variables has introduced additional 
terms in the exponent of the formal expression,
which are at most linear in $\dot{p}$ or $\dot{q}$, respectively. These terms
are not critical since, in the limit of diverging diffusion constant $\nu$, they 
will vanish. This means that the total change of the measure disappears
in the limit. Thus, one can write the path integral with the old measure
$d\mu_W^\nu$ instead of with the new $d\tilde{\mu}_W^\nu$:

\begin{equation}
\langle p^{\prime\prime}q^{\prime\prime}|e^{-i(RQ+SD)T}|p^\prime q^\prime
\rangle=
\lim_{\nu\rightarrow\infty}K_\nu {\pmb \int}^{p^{\prime\prime}, 
q^{\prime\prime}}_{p^\prime, q^\prime}
e^{-i{\textstyle \int}[q\,dp+(Rq+Spq)dt]}\:d\mu_W^\nu
\end{equation}

\noindent Now, the weak symbol can be read off:

\begin{equation}
h_w(p,q)=Rq+Spq
\end{equation}

The generalization to other Hamiltonians is based on the linearity, 
completeness, and irreducibility
of the basic operators $Q$ and $D$ by virtue of which $\lim_{J\rightarrow\infty}
\sum_{j=1}^J\alpha_je^{-i(R_jQ+S_jD)}$ weakly converges to any 
(bounded) operator such as $e^{-i{\cal H}T}$. Thus, 

\begin{eqnarray}
&&\langle p^{\prime\prime}q^{\prime\prime}|e^{-i{\cal H}T}|p^\prime q^\prime
\rangle=\lim_{J\rightarrow\infty}\langle p^{\prime\prime}q^{\prime\prime}|
\sum_{j=1}^J\alpha_je^{-i(R_jQ+S_jD)}|p^\prime q^\prime\rangle\nonumber\\
&=&\lim_{J\rightarrow\infty}\lim_{\nu\rightarrow\infty}K_\nu {\pmb \int}
e^{-i{\textstyle \int}q\,dp}[{\textstyle \sum}_{j=1}^J
\alpha_j e^{-i{\textstyle \int}(R_jq+S_jpq)dt}]\:d\mu_W^\nu
\end{eqnarray}

\noindent and the question, on which the next steps depend, is: can the
two limits be interchanged? In spite of some effort this question is not 
yet answered. 
Assuming that they can, however, one obtains

\begin{eqnarray}
&&\langle p^{\prime\prime}q^{\prime\prime}|e^{-i{\cal H}T}|p^\prime q^\prime
\rangle\nonumber\\
&=&\lim_{\nu\rightarrow\infty}K_\nu {\pmb \int}
e^{-i{\textstyle \int}q\,dp}[\lim_{J\rightarrow\infty}
{\textstyle \sum}_{j=1}^J
\alpha_j e^{-i{\textstyle \int}(R_jq+S_jpq)dt}]\:d\mu_W^\nu
\end{eqnarray}

\noindent The expression $[\lim_{J\rightarrow\infty}
{\textstyle \sum}_{j=1}^J
\alpha_j e^{-i{\textstyle \int}(R_jq+S_jpq)dt}]=:F[\int qdt, \int pqdt]$ 
is, unfortunately, not of
the form $e^{-i\int h_w(p,q)dt}$ for a general, local Hamiltonian $h_w$, e.g. 
$e^{-i\int q^2 dt}$ with Hamiltonian $q^2$. To produce local
Hamiltonians, one would need distributions $R(t)$ and $S(t)$ instead of the
constants $R$ and $S$. Then, taking e.g. $R(t)=\delta(t-\tau)$, 
one gets a local expression $q(\tau)$ and, by forming functions thereof, local
Hamiltonians. This was proposed in \cite{KlauderWCS}. However, the construction
of distributions from piecewise constant functions 
would require yet another limiting process, and, again, 
the interchangeability of the limits is questionable.



%
%

In the case of a linear Hamiltonian, the weak symbol was shown to be
$h_w(p,q)=Rq+Spq$. This is exactly what one would expect since the 
connection of the basic operators $Q$ and $D$ to classical variables is,
according to the weak correspondence principle,
$q$ and $pq$, respectively. But, the correspondence for a more general
Hamiltonian is not immediately clear and remains to be determined.

\subsection{Regularizing approach}

The idea for this second approach is the introduction of an additional 
regularization factor which will reintroduce a discrete ground state with
eigenvalue zero. Then, the construction of the path integral moves along the
same lines as in the coherent state case ($\beta>1/2$). The limit to 
remove the regularization is taken as the last step. 

For large $q$, the overlap $\langle pq|p^\prime q^\prime\rangle$ is 
proportional to $q^{-\beta}$. Because 
$0<\beta\leq1/2$, a regularization factor which is effective at 
infinity is required to 
produce Hilbert space vectors again.
Since, for $0<\beta\leq 1/4$, 
$\int_{-\infty}^{\infty}(c^2+p^2)^{-2\beta}dp=\infty$ (where $c$ is a constant), 
one 
must in this case regularize in $p$, too. For $1<4\beta<2$ this is not
required. A regularization in $p$ will make a regularization 
in $q$ (for small $q$) necessary as well \cite{Hartmann}.

\subsubsection*{Case ${\mathbf 1{\boldsymbol{/}}4{\boldsymbol{<}}
{\boldsymbol{\beta}}{\boldsymbol{\leq}}1{\boldsymbol{/}}2}$}

Let 

\begin{equation} 
\langle pq|rs\rangle_\varepsilon:=N_\varepsilon
\langle pq|rs\rangle e^{-(q+s)\varepsilon}\\
\end{equation}

\noindent be a normalized vector in $L^2(M_+)$ with normalization constant
$N_\varepsilon$. The extra factor $e^{-(q+s)\varepsilon}$ goes to one 
in the limit $\varepsilon\rightarrow0$. For arbitrary $x\in\mathbb{R}$,
$y\in\mathbb{R}^+$, the overlap $\langle xy|xy\rangle_\varepsilon$ equals
$N_\varepsilon
e^{-2y\varepsilon}$. Hence, one can write $\langle pq|rs\rangle
=\lim_{\varepsilon\rightarrow0}\langle xy|xy\rangle_\varepsilon^{-1}
\langle pq|rs\rangle_\varepsilon$ in a self-consistent way without explicitly
referring to the normalization constant. The following notation is used:
 
\begin{equation*}
\langle xy|xy\rangle_\varepsilon=:c_{\beta,\varepsilon}
\end{equation*}

The new operator $B_\varepsilon$, which annihilates the 
modified kernel, is derived by exploiting analyticity:
$[(q^{-1}+s^{-1})+i\beta^{-1}(p-r)]^{-2\beta}=:Y$ is analytic, so 
$\partial_{(q^{-1}-i\beta^{-1}p)}Y=\frac{1}{2}
(-q^2\partial_q+i\beta\partial_p)Y=0$. 
Write $Y$ as $e^{q\varepsilon}(qs)^{\beta}\langle pq|
rs\rangle_\varepsilon$, and move $e^{q\varepsilon}(qs)^{\beta}$ to the left of
this operator. Then, $e^{q\varepsilon}(qs)^{\beta}$ 
can be cancelled since the expression is everywhere 
non-zero. The result is the new operator
 
\begin{equation*}
B_\varepsilon=(\beta^{-1}q\partial_q
+\beta^{-1}q\varepsilon+1-iq^{-1}\partial_p)
\end{equation*}

\noindent for which $B_\varepsilon
\langle pq|rs\rangle_\varepsilon=0$.
Define $A_\varepsilon:=\frac{1}{2}\beta B_\varepsilon^\dagger B_\varepsilon$ 
then:

\begin{eqnarray}
A_\varepsilon
&=& {\textstyle\frac{1}{2}}\beta(-iq^{-1}\partial_p+1-\beta^{-1}\partial_qq+
\beta^{-1}q\varepsilon)(-iq^{-1}\partial_p+1+
\beta^{-1}q\partial_q+\beta^{-1}q\varepsilon)\nonumber\\
&=& {\textstyle\frac{1}{2}}\{\beta[-iq^{-1}\partial_p+1+\beta^{-1}q
\varepsilon]^2-\beta^{-1}\partial_qq^2\partial_q
-1-2\beta^{-1}q\varepsilon\}
\end{eqnarray}

\noindent $A_\varepsilon$ can be shown to be essentially self-adjoint  
since the deficiency index equation $[(A^\dagger_\varepsilon\pm i)\psi]
(p,q)=0$ has no solution \cite{Hartmann}. In a slight abuse of notation the
closure of this operator will be denoted by $A_\varepsilon$ as well. It
is a self-adjoint, non-negative operator with zero in its discrete spectrum.

The Feynman-Kac-Stratonovich representation of the kernel of the operator
$e^{-\nu TA_\varepsilon}$ is (see Appendix \ref{A} for the derivation)

\begin{eqnarray*}
&&e^{-\nu A_\varepsilon T}\delta(p-p^\prime)
\delta(q-q^\prime)|_{p=p^{\prime\prime}, q=q^{\prime\prime}}\nonumber\\
&=&e^{\nu T/2}\pmb{\int} e^{-i{\textstyle\int}(q+\beta^{-1}q^2
\varepsilon)dp
+\nu{\textstyle\int}\beta^{-1}q\varepsilon dt}\:d\mu_W^\nu
\end{eqnarray*}

\noindent and it follows that

\begin{eqnarray}
&&\langle p^{\prime\prime}q^{\prime\prime}|
p^\prime q^\prime\rangle=
\lim_{\varepsilon\rightarrow 0}c_{\beta,\varepsilon}^{-1}
\langle p^{\prime\prime}q^{\prime\prime}|
p^\prime q^\prime\rangle_\varepsilon\nonumber\\
&=&\lim_{\varepsilon\rightarrow 0}\lim_{\nu\rightarrow\infty}
c_{\beta,\varepsilon}^{-1}
e^{\nu T/2}\pmb{\int} e^{-i{\textstyle\int}(q+\beta^{-1}q^2
\varepsilon)dp
+\nu{\textstyle\int}\beta^{-1}q\varepsilon dt}\:d\mu_W^\nu
\end{eqnarray}

The stochastic processes involved are still Brownian bridges, and, 
when the stochastic integrals are interpreted in the Stratonovich sense,
canonical (coordinate) transformations can be made in the same way as before. 
Thus, the geometric nature of the quantization is
preserved.

\subsubsection*{Case ${\mathbf 0{\boldsymbol{<}}{\boldsymbol{\beta}}
{\boldsymbol{\leq}}1{\boldsymbol{/}}4}$}

For a parameter $\beta\leq1/4$, a regularization for large $q$ is not enough. It 
turns out that an additional $p$-regularization will even make a regularization
for small
$q$ necessary (otherwise the overlap would be square integrable, but not in the
domain of $A_\varepsilon$). 

In the present case, let

\begin{equation}
\langle pq|rs\rangle_\varepsilon:=N_\varepsilon
\langle pq|rs\rangle e^{-(q+s)\varepsilon
-(q^{-1}+s^{-1})\varepsilon-(p^2+r^2)\varepsilon}
\end{equation}

\noindent where $\langle pq|rs\rangle=(qs)^{-\beta}2^{2\beta}[(q^{-1}+s^{-1})
+i\beta^{-1}(p-r)]^{-2\beta}$ is the (weak coherent state) overlap which is
analytic in the complex variable $z:=q^{-1}+i\beta^{-1}p$, 
apart from the factor $(qs)^{-\beta}$. One can write the analytic part
(previously called $Y$) as $e^{(q+s)\varepsilon+(q^{-1}+s^{-1})
\varepsilon+(p^2+q^2)\varepsilon}(qs)^\beta\langle pq|rs\rangle_\varepsilon$, 
and let the differential operator $\partial_{q^{-1}-i\beta^{-1}p}
=\frac{1}{2}(-q^2\partial_q+i\beta\partial_p)$ act on this expression.
Using $\partial_{z^*}f=0$ (valid for an
analytic function), this results in the new operator

\begin{equation*}
B_\varepsilon=\beta^{-1}q\partial_q+1+\beta^{-1}q\varepsilon
+\beta^{-1}q^{-1}\varepsilon-2ipq^{-1}\varepsilon-iq^{-1}\partial_p
\end{equation*}

\noindent for which $B_\varepsilon\langle pq|rs\rangle_\varepsilon=0$. 
As before, define
$A_\varepsilon:=\frac{1}{2}\beta B^\dagger_\varepsilon B_\varepsilon$, then

\begin{eqnarray}
A_\varepsilon&=&{\textstyle\frac{1}{2}}
\{\beta(-iq^{-1}\partial_p+1+\beta^{-1}q\varepsilon
\beta^{-1}q^{-1}\varepsilon)^2-2\beta q^{-2}\varepsilon\nonumber\\
&&+4ip\partial_q\varepsilon+4\beta p^2q^{-2}
\varepsilon^2-\beta^{-1}\partial_1q^2\partial_q-1-2\beta^{-1}q\varepsilon\}
\end{eqnarray}

Instead of trying to solve
the deficiency index equation for the ``new" $A_\varepsilon$, one can avoid the
question about self-adjointness altogether.

Assume $A_\varepsilon$ is not self-adjoint. The (sesquilinear) form
$s_\varepsilon(x,y):=\langle x|A_\varepsilon y\rangle$ generated by 
$A_\varepsilon$ is closable since $A_\varepsilon$ is symmetric and bounded 
below
\cite{Blank}. There is a bijection between the set of all (densely defined) 
closed, below-bounded forms and the set of all self-adjoint, below-bounded 
operators. Let $\bar{s}_\varepsilon$ be the closure of the form generated by 
$A_\varepsilon$ and $A_{\bar{s}_\varepsilon}$ be the self-adjoint operator
associated with $\bar{s}_\varepsilon$. Then, $A_{\bar{s}_\varepsilon}$ 
preserves the lower
bound and is called the Friedrichs' extension of the operator $A_\varepsilon$. 
[It is the unique extension fulfilling $D(A_{\bar{s}_\varepsilon})\subset 
D(\bar{s}_\varepsilon)$ \cite{Blank}.]

In a slight abuse of notation $A_{\bar{s}_\varepsilon}$ will be written as
$A_\varepsilon$. So from now on, $A_\varepsilon$ denotes 
the Friedrichs' extension (which is
trivial in the case that $A_\varepsilon$ is already self adjoint). Then
it is clear that $A_\varepsilon$ is non-negative.

The Feynman-Kac-Stratonovich representation of the kernel of the 
operator
$\exp\{-\nu TA_\varepsilon\}$ is derived in much the same way as before
(see Appendix \ref{A}) 

\begin{eqnarray*}
&&e^{\nu T/2}e^{-i\beta^{-1}\varepsilon(p^{\prime\prime}-p^\prime)}
\nonumber\\
&&\times\pmb{\int} e^{-i{\textstyle\int}[
(q+\beta^{-1}q^2\varepsilon)\,dp-2\beta p q^{-2}\varepsilon\,dq]
+\nu{\textstyle\int}[\beta q^{-2}\varepsilon+\beta^{-1}q\varepsilon
] dt}\:d\mu_W^\nu
\end{eqnarray*}

\noindent Partial integration, i.e., $2\beta\varepsilon pq^{-2}dq
=-2\beta\varepsilon pd(q^{-1})=-2\beta\varepsilon pq^{-1}|^{(p^{\prime\prime},
q^{\prime\prime})}_{(p^\prime,q^\prime)}+2\beta\varepsilon q^{-1}dp$, leads to

\begin{eqnarray}\label{Kzero2}
&&e^{\nu T/2}e^{-i\beta^{-1}\varepsilon(p^{\prime\prime}-p^\prime)
-2i\beta\varepsilon(p^{\prime\prime}q^{\prime\prime-1}-p^\prime q^{\prime-1})}
\nonumber\\
&&\times\pmb{\int} e^{-i{\textstyle\int}
(q+\beta^{-1}q^2\varepsilon-2\beta q^{-1}\varepsilon)dp
+\nu{\textstyle\int}[\beta q^{-2}\varepsilon+\beta^{-1}q\varepsilon
] dt}\:d\mu_W^\nu
\end{eqnarray}

The phase factors in Eq. (\ref{Kzero2}) are $\nu$-independent, so
they come outside of the $\nu$-limit, where the $\varepsilon$-limit 
renders them unity. Finally, 
one gets:

\begin{eqnarray}\label{Kzero3}
&&\langle p^{\prime\prime}q^{\prime\prime}|p^\prime q^\prime\rangle
=\lim_{\varepsilon\rightarrow0}\lim_{\nu\rightarrow\infty}
c_{\beta,\varepsilon}^{-1}
\langle p^{\prime\prime}q^{\prime\prime}|p^\prime q^\prime\rangle_\varepsilon
\nonumber\\
&:=&\lim_{\varepsilon\rightarrow0}\lim_{\nu\rightarrow\infty}
c_{\beta,\varepsilon}^{-1}
e^{\nu T/2}\nonumber\\
&&\times
\pmb{\int} e^{-i{\textstyle\int}
(q+\beta^{-1}q^2\varepsilon-2\beta q^{-1}\varepsilon)dp
+\nu{\textstyle\int}[\beta q^{-2}\varepsilon+\beta^{-1}q\varepsilon
] dt}\:d\mu_W^\nu
\end{eqnarray}

\noindent This is the path integral representation for $0<\beta\leq1/4$.

\subsection*{Introducing dynamics}

Dynamics is introduced in the same way as for the spectral approach. For a 
linear Hamiltonian ${\cal H}=RQ+SD$, the problem is already solved as
it reduces to an overlap with modified ending points. What remains to do is
to write down the path integral. This is straightforward since everything
stated previously concerning the measure, etc., remains valid and the formula
for $1/4<\beta\leq1/2$ is

\begin{eqnarray}
&& \langle p^{\prime\prime}q^{\prime\prime}|e^{-i(RQ+SD)T}|p^\prime q^\prime
\rangle
=\langle p^{\prime\prime}e^{ST}+{\textstyle\frac{R}{S}}
(e^{ST}-1),q^{\prime\prime}e^{-ST}|p^\prime q^\prime\rangle\nonumber\\
&=&\lim_{\varepsilon\rightarrow0}\lim_{\nu\rightarrow\infty}
c_{\beta,\varepsilon}^{-1}e^{\nu T/2}
{\pmb \int}_{p^\prime,q^\prime}^{p^{\prime\prime}e^{ST}+
{\textstyle\frac{R}{S}}(e^{ST}-1),q^{\prime\prime}e^{-ST}}\!\!\!\!\!\!\!\exp\{-i
{\textstyle\int}(q+\beta^{-1}q^2\varepsilon)dp\nonumber\\
&&\hspace{7.8cm}
+\nu{\textstyle\int}
\beta^{-1}q\varepsilon dt\}\:d\mu_W^\nu\nonumber\\
&=&\lim_{\varepsilon\rightarrow0}\lim_{\nu\rightarrow\infty}
c_{\beta,\varepsilon}^{-1}e^{\nu T/2}
{\pmb \int}_{p^\prime,q^\prime}^{p^{\prime\prime},q^{\prime\prime}}
\exp\{-i
{\textstyle\int}(qe^{-St}+\beta^{-1}q^2e^{-2St}\varepsilon)\nonumber\\
&&\times d[pe^{St}+
{\textstyle\frac{R}{S}}(e^{St}-1)]
+\nu{\textstyle\int}
\beta^{-1}qe^{-St}\varepsilon dt\}\:d\mu_W^\nu\nonumber\\
&=&\lim_{\varepsilon\rightarrow0}\lim_{\nu\rightarrow\infty}
c_{\beta,\varepsilon}^{-1}e^{\nu T/2}
{\pmb \int}_{p^\prime,q^\prime}^{p^{\prime\prime},q^{\prime\prime}}
\exp\{-i[{\textstyle\int}(q+\beta^{-1}q^2e^{-St}\varepsilon)dp\nonumber\\
&&+{\textstyle\int}(q+\beta^{-1}q^2e^{-St}\varepsilon)(Sp+R)dt]
+\nu{\textstyle\int}
\beta^{-1}qe^{-St}\varepsilon dt\}\:d\mu_W^\nu
\end{eqnarray}

\noindent Introducing
the new variable $q_\varepsilon:=q+\beta^{-1}q^2e^{-St}\varepsilon$, 
and the new measure
$d\mu_W^{\nu,\varepsilon}:=\exp\{\nu{\textstyle\int}
\beta^{-1}qe^{-St}\varepsilon dt\}\:d\mu_W^\nu$, 
the complexity of the final expression can be hidden. The new measure is 
equivalent to the
old Wiener measure because the factor $\exp\{\nu{\textstyle\int}
\beta^{-1}qe^{-St}\varepsilon dt\}$ serves as a Radon-Nykodym derivative. 
Then, the 
formula resembles the path integral for coherent states and reads

\begin{equation}
\lim_{\varepsilon\rightarrow0}\lim_{\nu\rightarrow\infty}
c_{\beta,\varepsilon}^{-1}e^{\nu T/2}
{\pmb \int}_{p^\prime,q^\prime}^{p^{\prime\prime},q^{\prime\prime}}
e^{-i[{\textstyle\int}q_\varepsilon dp
+(Spq_\varepsilon+Rq_\varepsilon) dt]}\:d\mu_W^{\nu,\varepsilon}
\end{equation}

\noindent The $\varepsilon$-modified Hamiltonian is given by the weak modified 
symbol
$h_{w,\varepsilon}:=Rq_\varepsilon+Spq_\varepsilon$.

The same procedure for $0<\beta\leq1/4$ leads to:

\begin{eqnarray}
&& \langle p^{\prime\prime}q^{\prime\prime}|e^{-i(RQ+SD)T}|p^\prime q^\prime
\rangle\nonumber\\
&=&\lim_{\varepsilon\rightarrow0}\lim_{\nu\rightarrow\infty}
c_{\beta,\varepsilon}^{-1}e^{\nu T/2}
{\pmb \int}_{p^\prime,q^\prime}^{p^{\prime\prime},q^{\prime\prime}}
\exp\{-i[{\textstyle\int}(q+\beta^{-1}q^2e^{-St}\varepsilon
-2\beta q^{-1}e^{2St}\varepsilon)dp\nonumber\\
&&+{\textstyle\int}(q+\beta^{-1}q^2e^{-St}\varepsilon-2\beta q^{-1}e^{2St}
\varepsilon)
(Sp+R)dt]\nonumber\\
&&+\nu{\textstyle\int}
(\beta^{-1}qe^{-St}\varepsilon+\beta q^{-2}e^{2St}\varepsilon
) dt\}\:d\mu_W^\nu\nonumber\\
&=&\lim_{\varepsilon\rightarrow0}\lim_{\nu\rightarrow\infty}
c_{\beta,\varepsilon}^{-1}e^{\nu T/2}
{\pmb \int}_{p^\prime,q^\prime}^{p^{\prime\prime},q^{\prime\prime}}\!\!\!\!
\exp\{-i[{\textstyle\int}\tilde{q}_\varepsilon dp+{\textstyle\int}
(Sp\tilde{q}_\varepsilon+R\tilde{q}_\varepsilon) dt]\}\,
d\tilde{\mu}_W^{\nu,\varepsilon}
\end{eqnarray}

\noindent Here, the variable $\tilde{q}_\varepsilon:=
(q+\beta^{-1}q^2e^{-St}\varepsilon-2\beta q^{-1}e^{2St}\varepsilon)$ and
the Radon-Nykodym measure $d\tilde{\mu}_W^{\nu,\varepsilon}:=
\exp\{\nu{\textstyle\int}
(\beta^{-1}qe^{-St}\varepsilon+\beta q^{-2}e^{2St}\varepsilon
) dt\}\:d\mu_W^\nu$ were used. 
The weak modified
symbol is now $h_{w,\varepsilon}=R\tilde{q}_\varepsilon
+Sp\tilde{q}_\varepsilon$.

The problem of 
how this can be extended to, say, all polynomial Hamiltonians was
already discussed in the spectral approach. 
Here, on the other hand, there could be a second possibility to proceed. With 
the discrete ground state artifically reintroduced, it seems possible to 
construct the path integral in essentially the same way as for zero Hamiltonian.
The operator $\nu A_\varepsilon$ has to be replaced by $\nu A_\varepsilon
+ih_{w,\varepsilon}$, and the conditions required for the construction will
imply restrictions for the functions $h_{w,\varepsilon}$ (see \cite{DKP} for
a guideline to the proof). Observe, that this
weak modified symbol does not necessarily have to be the same as the one
mentioned in the previous parts of the subsection.


\section*{Acknowledgments}

Thanks are expressed to Wayne Bomstad for helpful remarks. 
One of the authors, Lorenz
Hartmann, would like to thank Prof. Dr. A. Schenzle for his support and 
feels indebted to the University of Florida for its hospitality.\\

\section*{Appendix A: Feynman-Kac-Stratonovich \\representation 
of the operator $A_\varepsilon$}\label{A}

{\bf The case $1/4<\beta\leq 1/2$}\\

The Feynman-Kac-Stratonovich representation of the kernel of 
the operator $\exp\{-\nu TA_\varepsilon\}$ is derived in the following
way:

{\allowdisplaybreaks
\begin{eqnarray*}
&&\exp\{-\nu TA_\varepsilon\}\delta(p-p^\prime)
\delta(q-q^\prime)|_{p=p^{\prime\prime},q=q^{\prime\prime}}\\
&=&\exp\{-{\textstyle\frac{1}{2}}\nu T
[\beta(-iq^{-1}\partial_p+1+\beta^{-1}q\varepsilon)^2-\beta^{-1}
\partial_qq^2\partial_q-1-2\beta^{-1}q\varepsilon]\}\\
&&\times\int e^{ix(p-p^\prime)-ik(q-q^\prime)}
{\textstyle\frac{dxdk}{(2\pi)^2}}|_{p=p^{\prime\prime},q=q^{\prime\prime}}\\
&=&e^{\nu T/2}\lim_{N\rightarrow\infty}[\exp\{-
{\textstyle\frac{1}{2}}\nu\delta[\beta(-i q^{-1}
\partial_p+1+\beta^{-1}q\varepsilon)^2-2\beta^{-1}q\varepsilon]\}\\
&&\times \exp\{-
{\textstyle\frac{1}{2}}\nu\delta(-\beta^{-1}\partial_qq^2\partial_q)\}]^N
\int e^{ix(p-p^\prime)-ik(q-q^\prime)}
{\textstyle\frac{dxdk}{(2\pi)^2}}|_{p=p^{\prime\prime},q=q^{\prime\prime}}\\
&=&\lim_{N\rightarrow\infty}e^{\nu T/2}
\int \exp\{i\sum x_{l+1/2}(p_{l+1}-p_l)-ik_{l+1/2}(q_{l+1}-q_l)\}\\
&&\times\exp\{-{\textstyle\frac{1}{2}}\nu\delta\sum
[\beta(q_l^{-1}x_{l+1/2}+1+\beta^{-1}q_l\varepsilon)^2-2\beta^{-1}q_l
\varepsilon]\}\\
&&\times \exp\{-{\textstyle\frac{1}{2}}
\nu\delta\sum\beta^{-1}k_{l+1/2}^2q_l^2\}\prod_{l=0}^N
\frac{dk_{l+1/2}dx_{l+1/2}}{(2\pi)^2}\prod_{l=1}^Ndp_ldq_l\\
&=:&e^{\nu T/2}{\cal N}\int \exp\{i\int(x\dot{p}-k\dot{q})dt\}\\
&&\times\exp\{-{\textstyle\frac{1}{2}}\nu\int
\{\beta(q^{-1}x+1+\beta^{-1}q\varepsilon)^2-2\beta^{-1}q\varepsilon+
\beta^{-1}k^2q^2\}dt\}\\
&&\times{\cal D}x{\cal D}k{\cal D}p{\cal D}q\\
&=&e^{\nu T/2}{\cal N}\int\exp\{i\int[(x-q-\beta^{-1}q^2\varepsilon)\dot{p}
-k\dot{q}]\}\\
&&\times\exp\{-{\textstyle\frac{1}{2}}
\nu\int(\beta q^{-2}x^2-2\beta^{-1}q\varepsilon
+\beta^{-1}k^2q^2)dt\}{\cal D}x{\cal D}k{\cal D}p{\cal D}q\\
&=&e^{\nu T/2}{\cal N}\int\exp\{-i\int(q+\beta^{-1}q^2
\varepsilon)\dot{p}dt\}\\
&&\times\exp\{{\textstyle\frac{1}{2}}\nu\int2\beta^{-1}q\varepsilon dt\}
\exp\{-{\textstyle\frac{1}{2\nu}}
\int[\beta^{-1}q^2\dot{p}^2+\beta q^{-2}\dot{q}^2]dt\}
{\cal D}p{\cal D}q
\end{eqnarray*}

\noindent with $N=T/\delta$. The Lie-Trotter product formula was used to go from
the second to the third equality. The indices $l+1/2$ and $l$ serve to
emphasize that
the temporal lattice points must not coincide for $x$,$p$ or $q$,$k$, 
respectively.
(This would violate the Heisenberg uncertainty principle.) For the
endpoints, the definitions
$p_0:=p^\prime$, $p_{N+1}:=p^{\prime\prime}$, $q_0:=q^\prime$ and
$q_{N+1}:=q^{\prime\prime}$ were made. Note that
$\exp\{\frac{1}{2}\nu\delta(-\beta^{-1}\partial_qq^2\partial_q)\}
\exp\{-ik(q-q^\prime)\}\approx \exp\{-\frac{1}{2}\nu\delta\beta^{-1}k^2q^2\}
\exp\{-ik(q-q^\prime)\}$ only to first order in $\delta$, but that is good 
enough for the path integral. In the second to 
last line, $x$ was substituted by 
$x+q-\beta^{-1}q^2\varepsilon$, and the $x$- and $p$-integrations were 
carried out.\\

{\bf The case $0<\beta\leq1/4$}\\

The Feynman-Kac-Stratonovich representation of the kernel of the 
operator
$\exp\{-\nu TA_\varepsilon\}$ is derived in much the same way as before, and, 
with
the same conventions for notation, it reads

\begin{eqnarray*}
&&\exp\{-\nu TA_\varepsilon\}\delta(p-p^\prime)
\delta(q-q^\prime)|_{p=p^{\prime\prime},q=q^{\prime\prime}}\\
&=&\exp\{-\nu T/2[\beta(-iq^{-1}\partial_p+1+\beta^{-1}q\varepsilon
+\beta^{-1}q^{-1}\varepsilon)^2-2\beta q^{-2}\varepsilon\\
&&+4ip\partial_q\varepsilon+4\beta p^2q^{-2}
\varepsilon^2-\beta^{-1}\partial_qq^2\partial_q-1-2\beta^{-1}q\varepsilon]\}\\
&&\times\int \exp\{ix(p-p^\prime)-ik(q-q^\prime)\}
\frac{dxdk}{(2\pi)^2}|_{p=p^{\prime\prime},q=q^{\prime\prime}}\\
&=&e^{\nu T/2}\lim_{N\rightarrow\infty}\bigg[\exp\{-\nu\delta/2\\
&&[\beta(-i q^{-1}
\partial_p+1+\beta^{-1}q\varepsilon+\beta^{-1}q^{-1}\varepsilon)^2
-2\beta q^{-2}\varepsilon-2\beta^{-1}q\varepsilon]\}\\
&&\times \exp\{-\nu\delta/2(-\beta^{-1}\partial_qq^2\partial_q)\}
\exp\{-\nu\delta/2\cdot4ip\partial_q\varepsilon\}\\
&&\times\exp\{-\nu\delta/2\cdot4\beta p^2q^{-2}\varepsilon^2\}\bigg]^N\\
&&\times\int \exp\{ix(p-p^\prime)-ik(q-q^\prime)\}
\frac{dxdk}{(2\pi)^2}|_{p=p^{\prime\prime},q=q^{\prime\prime}}\\
&=&\lim_{N\rightarrow\infty}e^{\nu T/2}
\int \exp\{i\sum x_{l+1/2}(p_{l+1}-p_l)-ik_{l+1/2}(q_{l+1}-q_l)\}\\
&&\times\exp\{-\nu\delta/2\sum
[\beta(q_l^{-1}x_{l+1/2}+1+\beta^{-1}q_l\varepsilon
+\beta^{-1}q_l^{-1}\varepsilon)^2\\
&&-2\beta q_l^{-2}\varepsilon-2\beta^{-1}q_l
\varepsilon]\}\\
&&\times \exp\{-\nu\delta/2\sum\beta^{-1}k_{l+1/2}^2q_l^2\}
\exp\{-\nu\delta/2\sum 4ip_l(-ik_{l+1/2})\varepsilon\}\\
&&\times\exp\{-\nu\delta/2\sum 4\beta p_l^2q_l^{-2}\varepsilon^2\}
\prod_{l=0}^N
\frac{dk_{l+1/2}dx_{l+1/2}}{(2\pi)^2}\prod_{l=1}^Ndp_ldq_l\\
&=:&e^{\nu T/2}{\cal N}\int \exp\{i\int(x\dot{p}-k\dot{q})dt\}\\
&&\times\exp\{-\nu/2\int
\{\beta(q^{-1}x+1+\beta^{-1}q\varepsilon+\beta^{-1}q^{-1}\varepsilon)^2
-2\beta q^{-2}\varepsilon-2\beta^{-1}q\varepsilon\\
&&+\beta^{-1}k^2q^2 +4pk\varepsilon+4\beta p^2q^{-2}\varepsilon^2\}dt\}
{\cal D}x{\cal D}k{\cal D}p{\cal D}q\\
&=&e^{\nu T/2}{\cal N}\int\exp\{i\int[(x-q-\beta^{-1}q^2\varepsilon
-\beta^{-1}\varepsilon)\dot{p}
-k\dot{q}]dt\}\\
&&\times\exp\{-\nu/2\int(\beta q^{-2}x^2
-2\beta q^{-2}\varepsilon-2\beta^{-1}q\varepsilon
+\beta^{-1}k^2q^2
+4pk\varepsilon\\
&&+4\beta p^2q^{-2}\varepsilon^2)dt\}
{\cal D}x{\cal D}k{\cal D}p{\cal D}q\\
&=&e^{\nu T/2}{\cal N}\int\exp\{-i\int(q+\beta^{-1}\varepsilon
+\beta^{-1}q^2\varepsilon)\dot{p}dt+i\int2\beta pq^{-2}\varepsilon
\dot{q}dt\}\\*
&&\times\exp\{\nu/2\int(2\beta^{-1}q\varepsilon
+2\beta q^{-2}\varepsilon
) dt\}\\*
&&\times\exp\{-1/(2\nu)\int[\beta^{-1}q^2\dot{p}^2+\beta q^{-2}\dot{q}^2]dt\}
{\cal D}p{\cal D}q
\end{eqnarray*}
}



\begin{thebibliography}{99}

\bibitem{DK}
I. Daubechies, J.R. Klauder, \textit{Quantum-mechanical path integrals with
Wiener measure for all polynomial Hamiltonians. II}, J. Math. Phys. 26, No.9,
2239-2256, 1985

\bibitem{DKP}
I. Daubechies, J.R. Klauder, T. Paul, \textit{Wiener measures for path integrals
with affine kinematic variables}, J. Math. Phys 28, 85-102, 1986

\bibitem{KlauderQiG}
J.R. Klauder, \textit{Quantization Is Geometry, after All}, Annals of Physics
188, 120-141, 1988

\bibitem{KlauderUQ}
J.R. Klauder, \textit{Understanding Quantization}, Foundations of Physics 27,
No. 11, 1467-1483, 1997

\bibitem{KlauderWCS}
J.R. Klauder, \textit{Coherent State Path Integral without Resolutions 
of Unity}, Foundations of Physics 31, No. 1, 57-67, 2001

\bibitem{Maraner}
P. Maraner, \textit{Landau Ground State on Riemannian Surface}, Modern Physics
Letters A, Vol. 7, No. 27, 2555-2558, 1992

\bibitem{KlauderAQG1}
J.R. Klauder, \textit{Noncanonical quantization of gravity. I. Foundations of
affine quantum gravity}, J. Math. Phys. 40, No. 11, 1999

\bibitem{KlauderAQG2}
J.R. Klauder, \textit{Noncanonical quantization of gravity. II. Constraints
and the physical Hilbert space}, J. Math. Phys. 42, No. 9, 2001

\bibitem{Aslaksen}
E.W. Aslaksen, J.R. Klauder, \textit{Continuous Representation Theory Using
the Affine Group}, J. Math. Phys. 10, 2267-2275, 1969

\bibitem{Klauder}
J.R. Klauder, B-S. Skagerstam, \textit{Coherent States, Applications in 
Physics and Mathematical Physics}, World Scientific, Singapore, 1985

\bibitem{Hartmann}
L. Hartmann, \textit{Weak Coherent States and their Path Integrals}, 
Diplom-thesis, University of Munich, August, 2003; hep-th/0308185

\bibitem{Grosche}
C. Grosche, \textit{The Path Integral on the
Poincar\'{e} Upper Half-Plane with a Magnetic Field and for the 
Morse Potential}, Annals of Physics 187, 110-134, 1988

\bibitem{Blank}
J. Blank, P. Exner, M. Havl\'{i}\v{c}ek, \textit{Hilbert Space Operators in
Quantum Physics}, AIP Series in Computational and Applied Mathematical Physics,
New York, 1994

\end{thebibliography}
\end{document}